\documentclass{aa}
\usepackage{psfig}
\def\mum{\hbox{\,$\mu$m}}
\newcommand{\etal}{et al.~}
\newcommand{\eg}{e.g.~}
\newcommand{\ie}{{\it{ie.~}}}

\newcommand{\HII}{\mbox{H\,{\sc ii}}}
\newcommand{\NeII}{\mbox{Ne\,{\sc ii}}}
\newcommand{\NeIII}{\mbox{Ne\,{\sc iii}}}
\newcommand{\ArII}{\mbox{Ar\,{\sc ii}}}
\newcommand{\ArIII}{\mbox{Ar\,{\sc iii}}}
\newcommand{\ArIV}{\mbox{Ar\,{\sc iv}}}
\newcommand{\OII}{\mbox{O\,{\sc ii}}}
\newcommand{\OIII}{\mbox{O\,{\sc iii}}}
\newcommand{\SII}{\mbox{S\,{\sc ii}}}

\newcommand{\NII}{\mbox{N\,{\sc ii}}}
\newcommand{\flux}{{\hbox{\,erg\,\,s$^{-1}$\,cm$^{-2}$\,sr$^{-1}$}}}

\begin{document}
\thesaurus{08	
	(09.09.1 Orion nebula; 09.09.1 Orion bar; 08.09.2 $\theta^2$~Ori~A;
	09.08.1; 09.04.1; 13.09.4)}
\title {Silicate emission in Orion
\thanks{Based on observations with ISO, an ESA project with instruments
funded by ESA member states (especially the PI countries: France,
Germany, the Netherlands and the United Kingdom) and with the
participation of ISAS and NASA.}}

\author{D. Cesarsky\inst{1,2}\and
        A.P. Jones\inst{1}\and
	J. Lequeux\inst{3}\and
	L. Verstraete\inst{1}}
\offprints{james.lequeux@obspm.fr}
\institute{	
Institut d'Astrophysique Spatiale, Bat. 121, Universit\'e Paris
XI, F-91405 Orsay CEDEX, France; 
{\it ant@ias.fr; verstra@ias.fr}
\and
Max Planck Institut f\"ur extraterrestrische Physik, D-85740 Garching,
Germany; {\it diego.cesarsky@mpe.mpg.de}
\and
DEMIRM,Observatoire de Paris, 61 Avenue de l'Observatoire, F-75014
Paris, France; {\it james.lequeux@obspm.fr}
}
\date{Received 7 December 1999; accepted 26 January 2000}
\maketitle
\begin{abstract}
We present mid--infrared spectro--imagery and high--resolution
spectroscopy of the Orion bar and of a region in the Orion
nebula. These observations have been obtained in the Guaranteed Time
with the Circular Variable Filters of the ISO camera (CAM-CVF) and
with the Short Wavelength Spectrometer (SWS), on board the European
Infrared Space Observatory (ISO). Our data shows emission from
amorphous silicate grains from the entire \HII~ region and around the
isolated O9.5V star $\theta^2$~Ori~A. The observed spectra can be
reproduced by a mixture of interstellar silicate and carbon grains
heated by the radiation of the hot stars present in the
region. Crystalline silicates are also observed in the Orion nebula
and suspected around $\theta^2$~Ori~A. They are probably of
interstellar origin.  The ionization structure and the distribution of
the carriers of the Aromatic Infrared Bands (AIBs) are briefly
discussed on the basis of the ISO observations.
\keywords	{ISM: Orion nebula		-
		ISM: Orion bar  		-
		ISM: \HII~ regions		-
		stars: $\theta^2$~Ori~A		-
		dust, extinction		-
		Infrared: ISM: lines and bands} 
\end{abstract}
\section{Introduction}
The Orion nebula is one of the most studied star--forming regions in
the Galaxy.  The ionizing stars of the Orion nebula (the Trapezium
stars, the hottest of which is $\theta^1$~Ori~C, O6) have eroded a
bowl--shaped \HII~ region into the surface of the Orion molecular
cloud. The Orion bar is the limb--brightened edge of this bowl where
an ionization front is progressing into the molecular cloud. It is
seen as an elongated structure at a position angle of approximately
60\degr. The Orion nebula extends to the North.  The Trapezium stars
are located at an angular distance of approximately 2.3 arc minutes
from the bar, corresponding to 0.35 pc at a distance of 500 pc.  The
molecular cloud extends to the other side of the bar, but also to the
back of the Orion nebula. The bright star $\theta^2$~Ori~A (O9.5Vpe)
lies near the bar, and is clearly in front of the molecular cloud
since its color excess is only E(B--V) $\simeq$ 0.2 mag.

\begin{figure*}[t!]
{\psfig{file=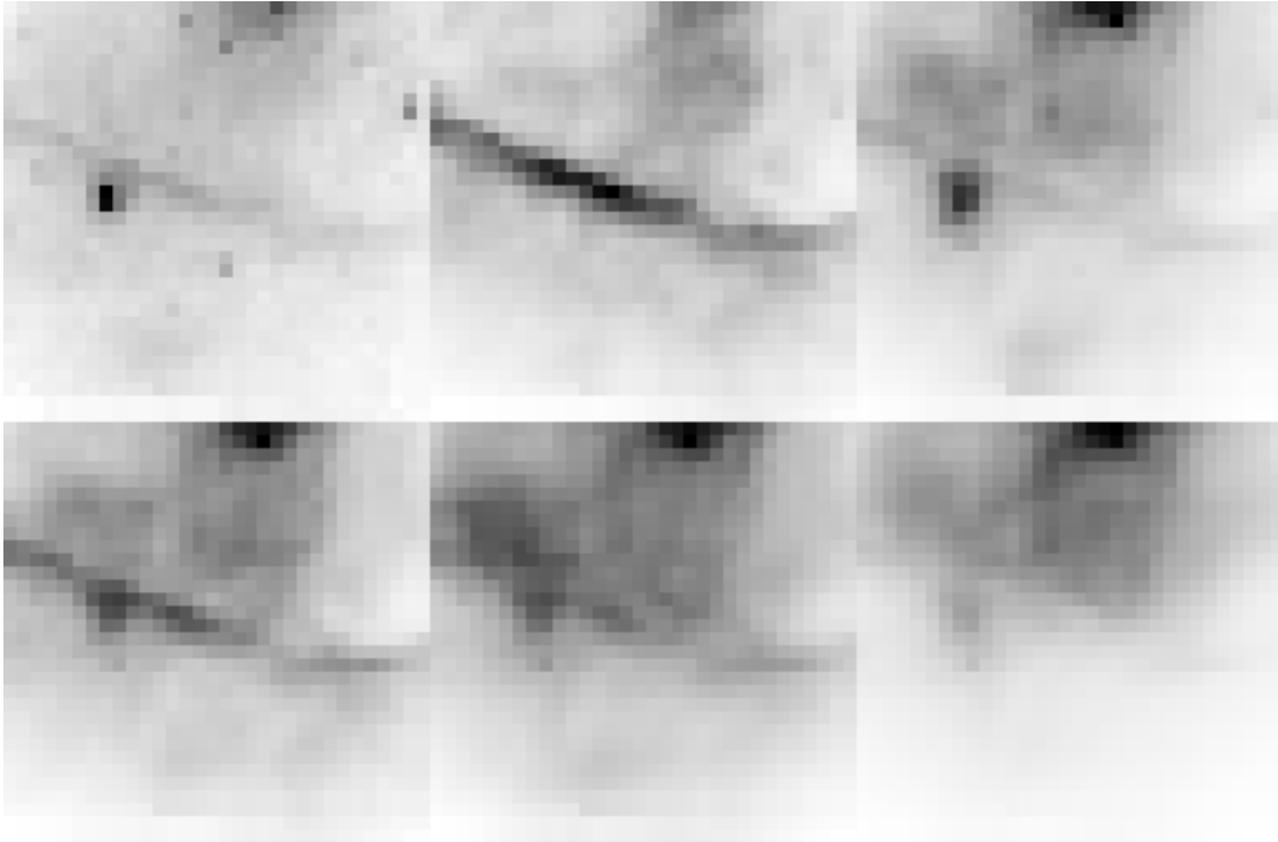,width=17cm,angle=00}}
\caption{Mosaic of six images of the Orion bar area (shown in
detector coordinates, a clockwise rotation of 10.4\degr~ is needed to
display the real sky orientation; see Fig. 2 for the equatorial
coordinates). Top row: {\it (left)} an image at 5.01\mum,
$\theta^2$~Ori~A is visible in the middle left of the image; {\it
(centre)} the Orion bar at the AIB wavelength of 6.2\mum; {\it
(right)} image at 9.5\mum (the wavelength of one of the silicate
features), note that $\theta^2$~Ori~A is again visible. Bottom row:
{\it (left)} image at the AIB wavelength of 11.3\mum~ and {\it
(centre)} image at 12.7\mum~(ISOCAM's CVF resolution blends [\NeII]
and the 12.7\mum~AIB feature); image at 15.6\mum {\it (right)},
wavelength of the [\NeIII] forbidden line.  }
\end{figure*}

Figs. 1 and 2 illustrate the geometry of the region observed. Fig. 1
shows six representative images of the region of the Orion bar (see
the figure caption for details). Fig. 2 shows the contours of the the
[\NeIII] 15.5\mum~ fine-structure line emission which delineates the
\HII~ region. The emission in one of the mid-IR bands (hereafter
called the Aromatic Infrared Bands, AIBs) at 6.2\mum~ traces the Orion
bar (an edge-on PhotoDissociation Region or PDR). The AIBs are usually
strongly emitted by PDRs.  The Trapezium region was avoided because of
possible detector saturation.

Pioneering infrared observations by Stein \& Gillet (\cite{Stein}) and
Ney \etal (\cite{Ney}) discovered interstellar silicate emission near
10\mum~ in the direction of the Trapezium. This was confirmed by
Becklin \etal (\cite{Becklin}) who also noticed extended silicate
emission around $\theta^2$~Ori~A. Since that time, interstellar
silicate emission has been found by the Infrared Space Observatory
(ISO) in the \HII~ region N 66 of the Small Magellanic Cloud (Contursi
\etal in preparation) and in a few Galactic compact \HII~ regions (Cox
\etal in preparation) and Photodissociation Regions (PDRs, Jones \etal
in preparation). The emission consists of two broad bands centered at
9.7 and 18\mum, which show little structure and are clearly dominated
by amorphous silicates.

We report in the present article ISO observations of the Orion bar and
of a part of the Orion nebula made with the Circular Variable Filter
of the ISO camera (CAM-CVF) which allowed imaging spectrophotometry of
a field $3\arcmin\times3\arcmin$ at low wavelength resolution (R
$\simeq$ 40). We also use an ISO Short--Wavelength Spectrometer (SWS)
observation which provides higher--resolution ($R\simeq 1000$)
spectroscopy at a position within the \HII~ region (see Fig. 2). This
spectrum was taken as part of the MPEWARM guaranteed-time program. We
show here that these new ISO data confirm and extend previous
observations of the amorphous silicate emission and also give evidence
for emission by crystalline silicates.

Sect. 2 of this paper describes the observations and data
reduction. In Sect. 3, we discuss the emission of dust and gas. The
silicate emission is characterized in section 4 through modelling of
the observed continuum IR emission. Finally, conclusions are presented
in Sect. 5. Our observations also give information on the
fine--structure lines and on the AIBs. This will be presented in
Appendices A and B respectively.

\section{Observations and data reduction}
Imaging spectrophotometry was performed with the 32$\times$32 element
mid-IR camera (CAM) on board the ISO satellite, using the Circular
Variable Filters (CVFs) (see Cesarsky et al. ~\cite{CCesarsky} for a
complete description). The observations employed the 6\arcsec~ per
pixel field-of-view of CAM. Full scans of the two CVFs in the
long-wave channel of the camera were performed with both increasing
and decreasing wavelength. The results of these two scans are almost
identical, showing that the transient response of the detector was
only a minor problem for these observations.  The total wavelength
range covered is 5.15 to 16.5\mum~ and the wavelength resolution
$\lambda/\Delta\lambda \simeq 40$.  10$\times$0.28 s exposures were
added for each step of the CVF, and 7 more at the first step in order
to limit the effect of the transient response of the detectors. The
total observing time was about 1 hour. The raw data were processed as
described in Cesarsky \etal~(\cite{M17}), with improvements described
by Starck \etal (\cite{Starck}) using the CIA software\footnote{CIA is
a joint development by the ESA Astrophysics Division and the ISOCAM
Consortium led by the ISOCAM PI, C. Cesarsky, Direction des Sciences
de la Mati\`ere, C.E.A., France}. The new transient correction
described by Coulais \& Abergel (\cite{Coulais}) has been applied but
the corrections introduced are minimal, as mentioned above. The bright
star $\theta^2$~Ori~A is visible in the maps of several spectral
components and has been used to re--position the data cube. This
involved a shift of only 2\arcsec. The final positions are likely to
be good to 3\arcsec~ (half a pixel).

\begin{figure} 
\vspace{-1cm}
{\psfig{file=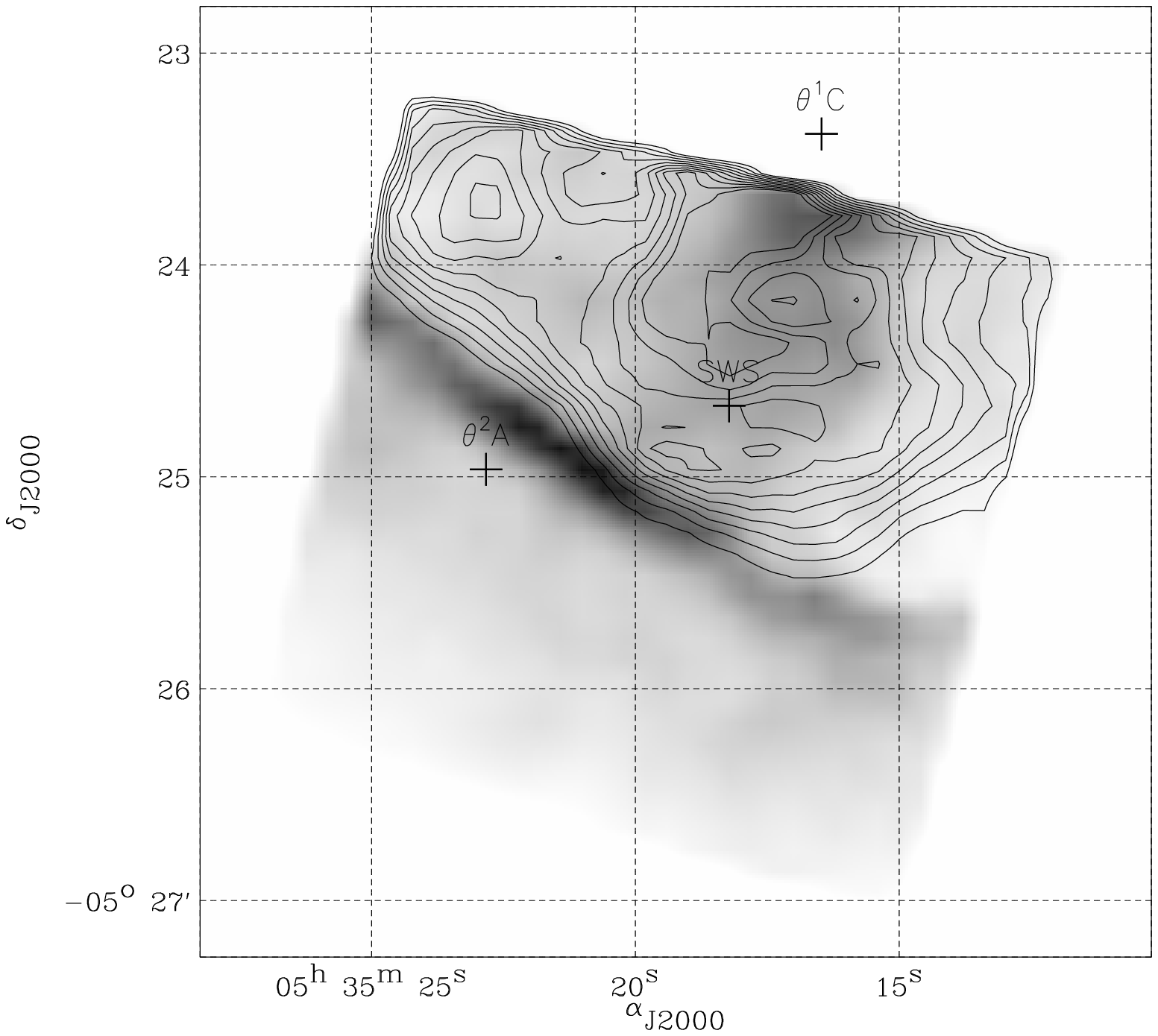,width=8.8cm,angle=00}}
\caption{[\NeIII] map (contours) and 6.2\mum~ Aromatic Infrared Band 
(AIB) map (grey scale). The grey scale corresponds to line intensities
from 0.01 to 0.30 \flux. The contours correspond to band intensities
of 0.015 to 0.06 \flux~ by steps of 0.05. The cross is at the position
of the O9.5V star $\theta^2$~Ori~A. The position of the hottest of the
Trapezium stars ($\theta^1$~Ori~C) is also indicated, outside the
observed field. The position of the SWS aperture in the direction of
the \HII~ region is shown by a black cross.  }
\end{figure}

All the maps presented here were obtained from the CVF data cube and
have approximately the same resolution: namely, 6\arcsec~ pixels at
the short wavelengths increasing to about 8\arcsec~ at 15\mum; see
Appendix C for more details.

In several of these maps a faint emission can be seen on the
south--east part of the ISOCAM field of view. This feature does not
correspond to anything conspicuous in published images of the region,
in particular in the near--IR images of Marconi \etal
(\cite{Marconi}). It is a spurious feature due to multiple reflections
of the strong Trapezium between the detector and the CVF filter wheel,
as shown by the ISOCAM ray tracing studies of Okumura (\cite{koko}).

The complete SWS scan (2.4-46\mum) was reduced with the latest version
of SWS-IA running at the Institut d'Astrophysique
Spatiale. Calibration files version CAL-030 were used.

Fig. 3 presents the SWS spectrum ($\lambda/\Delta\lambda \simeq 1000$)
obtained inside the \HII~ region at the position indicated on Fig. 2. Fig. 3
also shows the comparison of the SWS spectrum with that of the CAM-CVF pixels
averaged in the SWS aperture. The agreement between these spectra is excellent,
well within 20 percent for the continuum.

\section{Gas and dust emission}

The spectra towards the \HII~ region of the Orion nebula are shown in
Figs. 3 and 4.  The CAM-CVF spectrum of Fig. 3 is representative of
the whole field because CVF spectra obtained at different positions in
the \HII~ region and around $\theta^2$~Ori~A look qualitatively
similar (compare Figs. 3 and 10 which show the CVF spectra of
different pixels; note particularly the rising long wavelength portion
of the spectra). \\

In the SWS spectrum, a large number of unresolved lines from atoms,
ions and molecules are visible. We note the Pf$\alpha$ recombination
line of hydrogen (emitted by the warm, ionized gas of the \HII~
region) and the molecular hydrogen pure rotation lines S(2) and very
faintly S(3) and S(5) (stemming from the cooler, molecular PDR gas).
The simultaneous presence of these lines reflects the variety of
physical conditions present along the line of sight. Clearly, we are
looking at emission from the \HII~ region mixed with some emission
from the background PDR. These unresolved lines are briefly discussed
in Appendix A. \\

\begin{figure}[t!]
\vspace{-0.5cm}
{\psfig{file=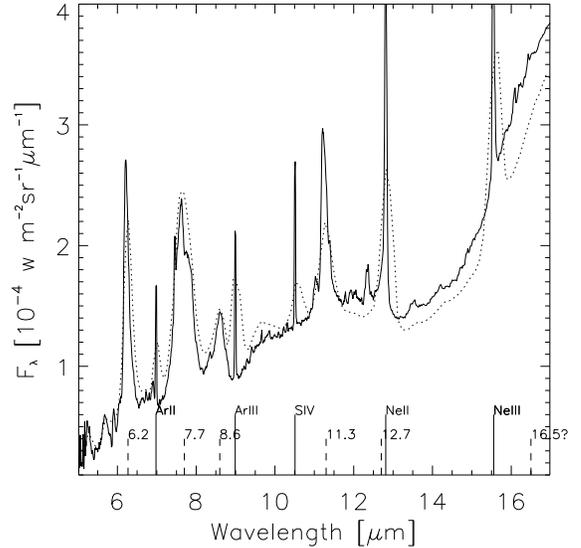,width=8.8cm,angle=00}}
\caption{The SWS spectrum (full line) compared to the CAM-CVF spectrum
(dotted line). In this latter case, all the CAM pixels falling in the
SWS aperture have been co-added.  }
\end{figure}

\begin{figure}[!t]
\vspace{0.3cm}
{\psfig{figure={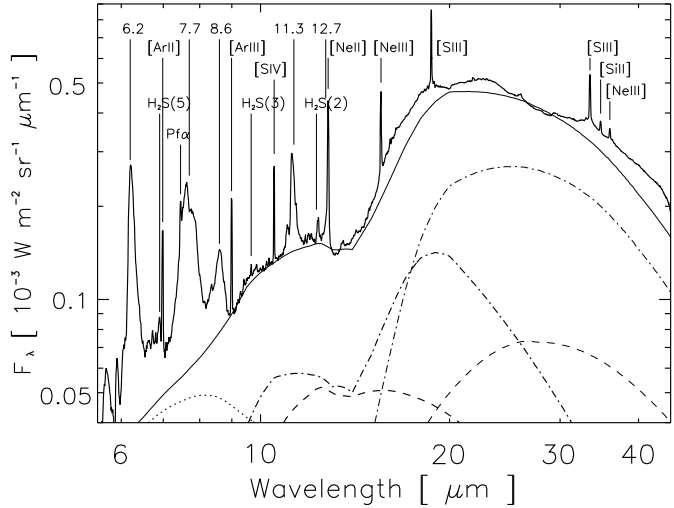},width=8.8cm } }
\caption{SWS spectrum in the Orion nebula at the position shown in
Fig. 2.  A fit to the spectrum (see Sect. 3 for details) is shown
which uses amorphous astronomical silicate (130~K: bold dashed-dotted,
and 80~K: light dashed-dotted), amorphous carbon (155~K: bold dashed,
and 85~K: light dashed), and amorphous carbon VSGs (300~K: dotted).
The total calculated spectrum is given by the thin solid line.  The
identification of the strongest spectral features is indicated.  }
\end{figure}

The other striking fact of the SWS spectrum is the strong continuum
peaking at about 25\mum. It is emitted by warm dust in the \HII~
region, but dust from the background PDR probably also
contributes. The broad emission bands of amorphous silicates centered
at 10 and 18\mum~ are visible. The classical AIBs at 6.2, 7.7, 8.6,
11.3 and 12.7\mum~ dominate the mid-IR part of the spectrum. As
discussed by Boulanger \etal \cite{bbcr}, the mid-IR spectrum can be
decomposed into Lorentz profiles (the AIBs) and an underlying
polynomial continuum. Maps of the various AIBs constructed in this way
all show the same morphology originating mainly from the PDR gas in
the Orion bar (see Appendix B). We will hereafter use the 6.2\mum-band
as representative of the behaviour of the AIBs. \\

In Fig. 5 we compare the behaviour of the mid-IR continuum emission
and of the AIBs. Clearly, the AIB emission is concentrated in the
Orion bar whereas the 15.5\mum-continuum emission extends throughout
the whole CAM field and shows a local peak around $\theta^2$~Ori~A
(note that the mid-IR emission around this star is foreground because
the star lies in front of the nebula). The continuum emission,
however, appears to peak towards $\theta^1$~Ori~C, outside the region
observed with ISOCAM.

The contrast in the emission morphology between the bands and
continuum can be interpreted in terms of the photodestruction of the
AIB carriers in the hard UV-radiation field of the \HII~ region. The
AIB carriers must be efficiently destroyed while the larger grains are
much more resistant (\eg Allain \etal 1996). We detail the modelling
of the dust thermal emission in the next section.
  
\begin{figure}[t!]
\vspace{-1cm}
{\psfig{file=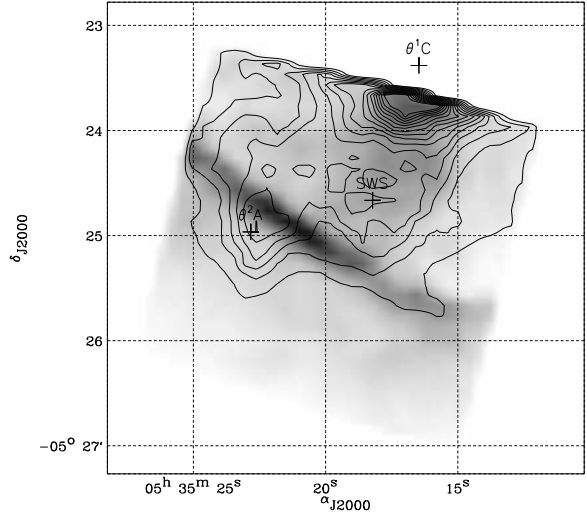,width=8.8cm,angle=00}}
\caption{Continuum emission at 15.5\mum~ (contours) superimposed on the
AIB 6.2\mum~ map (grey scale). The continuum flux was taken to be the
average of the flux on each side of the [\NeIII]15.5\mum~ line.  The
6.2\mum~feature strength was estimated as explained in the Appendix C.
The contours are from 10 to 80 Jy/pixel (1 pixel =
$6\arcsec\times6\arcsec$), by steps of 5 Jy/pixel; the grey scale map
spans 0.01 to 0.3 \flux. The position of $\theta^2$~Ori~A is indicated
by a cross.  }
\end{figure}

\subsection{Modelling the dust emission}

To account for the observed SWS spectra, we have calculated the
thermal equilibrium temperature of dust in the Orion \HII~ region as a
function of distance of the Orion bar from the Trapezium stars,
assuming that $\theta^1$ Ori C (an O6 star) dominates the local
radiation field. We use the optical constants of the amorphous
astronomical silicate of Draine (\cite{Draine85}) and of the amorphous
carbon AC1 of Rouleau \& Martin (\cite{Rouleau}). Assuming typical
interstellar grain sizes (e.g. Draine \& Lee \cite{DandL84}), we find
a temperature range of 85--145~K for amorphous silicates and a range
of 110--200~K for amorphous carbon, corresponding to grains of radius
1500 and 100 \AA~ respectively, at a distance of $\sim 0.35$~pc from
$\theta^1$ Ori C (the distance of the Orion Bar to the Trapezium
stars).

Using, for simplicity, discrete dust temperatures consistent with
those calculated above (T$_{\rm silicate}$ = 80~K and 130~K, T$_{\rm
carbon}$ = 85~K and 155~K) we are able to satisfactorily model the
continuum emission spectrum from the dust in the Orion \HII~ region at
the position of the ISO-SWS spectrum.  In Fig. 4 we show the
calculated emission spectrum from our model where we adopt the
carbon/silicate dust mass ratios of Draine \& Lee (\cite{DandL84}). In
the calculated spectrum we have included the emission from carbon
grains at 300~K, containing $\sim$ 1 percent.  of the total carbon
dust mass, in order to fit the short wavelength continuum
emission. The hot carbon grain emission mimics that of the
stochastically-heated Very Small Grains (VSGs, D\'esert \etal 1990).
The 300~K temperature represents a mean of the temperature
fluctuations for these small particles in the radiation field of
$\theta^1$ Ori C, and therefore indicates a lower mass limit of $\sim$
1 percent for the mass of the available carbon in VSGs.

The results of our model show that the emission feature in the 10\mum~
region is dominated by amorphous silicates at temperatures of the
order of 130~K, but that there may also be a small contribution from
amorphous carbon grains in the 12\mum~ region (Fig. 4). We also note
broad ``features'' in the SWS spectrum, above the modelled continuum
in Fig. 4, at $\sim 15-20$\mum, $\sim 20-28$\mum~ and longward of
32\mum, that are not explained by our model. These features bear a
resemblance to the major bands at 19.5, 23.7 and 33.6\mum~ seen in the
crystalline forsterite spectra of Koike \etal (\cite{Koike}) and of
Jaeger \etal (\cite{Jaeger}). Bands in these same wavelength regions
were noted by Jones \etal (\cite{Jones}) in the SWS spectra of the
M\,17 \HII~ region and were linked with the possible existence of
crystalline Mg-rich olivines in this object. Thus, similar broad
emission bands are now observed in the 15--40\mum~wavelength region of
the SWS spectra of two \HII~ regions (Orion and M\,17). These bands
resemble those of the crystalline Mg-rich silicate forsterite. Another
band at 9.6\mum~ is probably due to some sort of crystalline silicate,
and will be discussed in more details in the next section.

This dust model is simple--minded but emphasizes dust spectral
signatures in the mid-IR continuum which was the main aim here. More
detailed modelling treating temperature fluctuations and taking into
account the grain size distribution is underway (Jones \etal in
preparation).

%
The broad continua that lie above the model fit (i.e. $\sim 20-28$\mum~ and
$>$~32\mum, Fig. 4) can be associated with crystalline silicate emission bands.
This seems to be a robust conclusion of this study. The features are too narrow
to be explained by single-temperature blackbody emission and are therefore
likely to be due to blended emission features from different materials.
Unfortunately, having only one full SWS spectrum and CVF spectra that do not
extend beyong 18\mum, we are unable to say anything about the spatial variation
of these broad bands in the Orion region.

Interestingly, broad plateaux in the $\sim 15-20$\mum~ region have been
associated with large aromatic hydrocarbon species containing of the order of a
thousand carbon atoms (van Kerckhoven \etal \cite{vanKerckhoven}). However, in
this study the integrated intensity of the $\sim 15-20$\mum~ plateaux do vary
by a factor of up to 10 relative to the aromatic carbon features shortward of
13\mum.  Thus, the origin of these broad emission features does remain
something of an open question at this time.
%

\section{Tracing the silicate emission}

To delineate the spatial extent of the 10\mum-silicate emission
conspicuously visible in Fig. 3 and 4, we proceed as follows. We start
with the spectrum towards $\theta^2$~Ori~A, which shows the most
conspicuous silicate emission and we represent the AIBs by Lorentz
profiles, see Fig. 6 (top). Next we subtract them from the CVF
spectra. The remaining continuum has the generic shape of a blackbody
on top of which we see the broad bands corresponding to the silicate
emission, Fig. 6 (middle). Finally, we subtract a second order
polynomial from the continuum thus obtaining the well known silicate
emission profile at that position, Fig. 6 (bottom). The profile thus
obtained is then used as a scalable template to estimate the emission
elsewhere, see Appendix C for more details.

\begin{figure}[t!]
\vspace{-0.5cm}
{\psfig{file=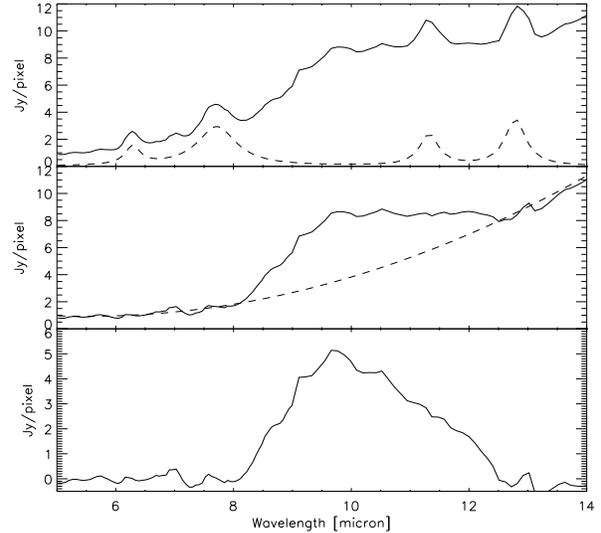,width=8.8cm,angle=00}}
\caption{{\it Top panel:} CVF spectrum towards $\theta^2$~Ori~A (solid
line).  The ordinates give fluxes in Jy per $6\arcsec\times6\arcsec$
pixel. A Lorentz fit to the AIBs is shown as the dotted line. {\it
Middle panel:} result of the subtraction of the AIBs from the CVF
spectrum. The fit to the continuum is shown by the dotted line. {\it
Bottom panel:} Residual from the middle figure, i.e. the suspected
amorphous silicate emission profile; notice the narrower bump near
9.6\mum.  }
\end{figure}

On top of the broad band of amorphous silicate centered near 9.7\mum~
we see a band centered at nearly 9.6\mum, which we ascribe to
crystalline silicates (Jaeger \etal \cite{Jaeger}). This band was also
used as a scalable template as explained above and in Appendix
C. Finally, the S(5) rotation line of H$_2$ at 6.91\mum~ is present
and is probably blended with the [\ArII] line at 6.99\mum.

In Fig. 7, we see that the spatial distribution of the 9.7\mum-feature
of amorphous silicate is quite similar to that of the
15.5\mum-continuum.  The 15.5\mum~ continuum emission includes a
strong contribution from silicates (see Fig. 4), but a peak in the
silicate emission around $\theta^2$~Ori~A is also evident. The
silicate emission is thus predominantly due to larger grains. The
narrower 9.6\mum~feature is mapped in Fig. 8. We note its similarity
to the distribution of the 9.7\mum~broad band: this fact lends support
to our assignation of this band to crystalline silicate. \\

\begin{figure}[t!]
\vspace{-1cm}
{\psfig{file=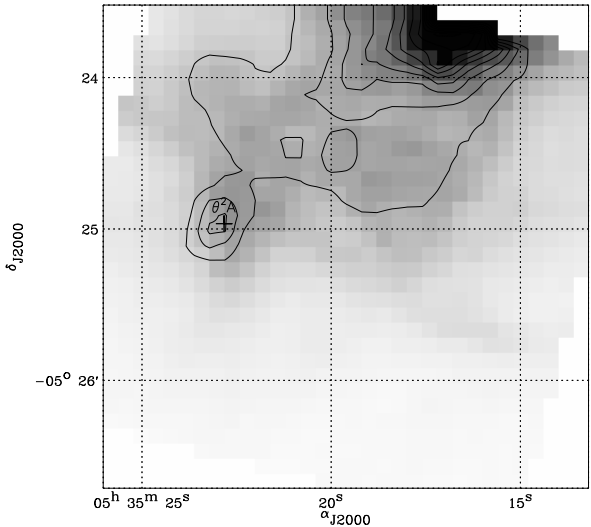,width=8.8cm,angle=00}}
\caption{Map of the intensity of the broad 9.7\mum~ band of amorphous 
silicates (contours) superimposed on the 15.5\mum~ continuum map (grey
scale).  Note the bright silicate emission around $\theta^2$~Ori~A
(cross). The contours correspond to integrated band intensities from
0.25 to 0.7 \flux~ by steps of 0.05; the gray image spans from 1 to 80
Jy/pixel.  }
\end{figure}

\begin{figure}[t!]
\vspace{-1cm}
{\psfig{file=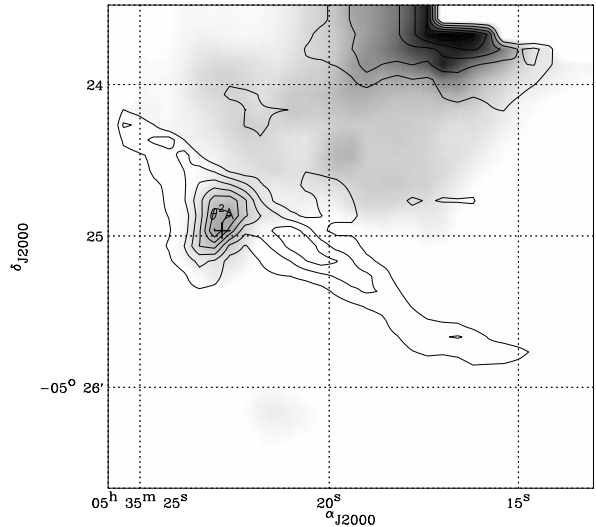,width=8.8cm,angle=00}}
\caption{Map of the 9.6 micron feature map (contours) superimposed on
the map of the broad 9.7\mum~ band of amorphous silicates (grey scale
spanning 0.1 to 10 \flux. The contours correspond to integrated band
intensities from 0.02 to 0.11 \flux~ by steps of 0.001. The shift with
respect to the position of $\theta^2$~Ori~A (cross) is by less than
one pixel and may not be significant.  }
\end{figure}

\begin{figure}[t!]
\vspace{-1cm}
{\psfig{file=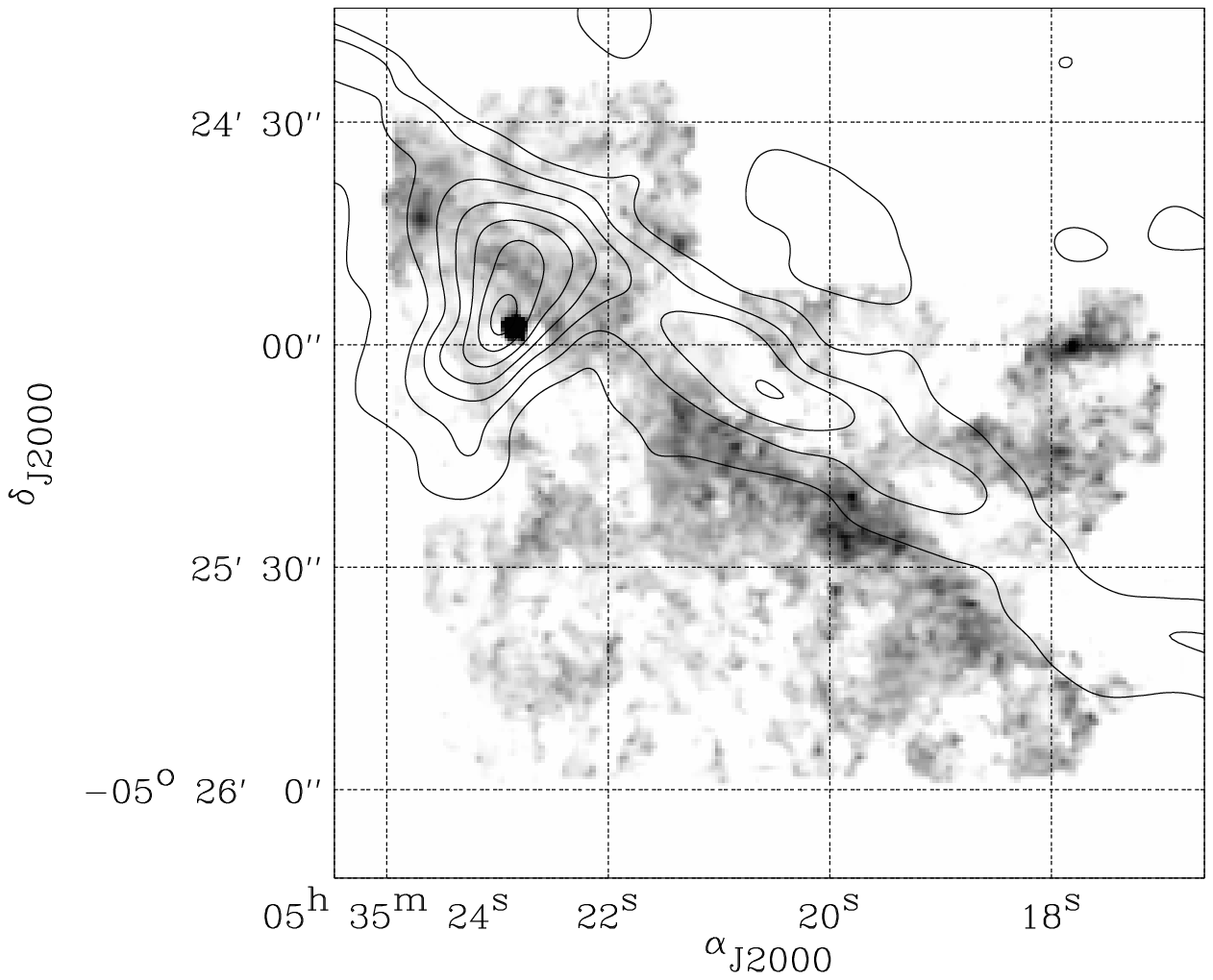,width=8.8cm,angle=00}}
\vspace{0.3cm}
{\psfig{file=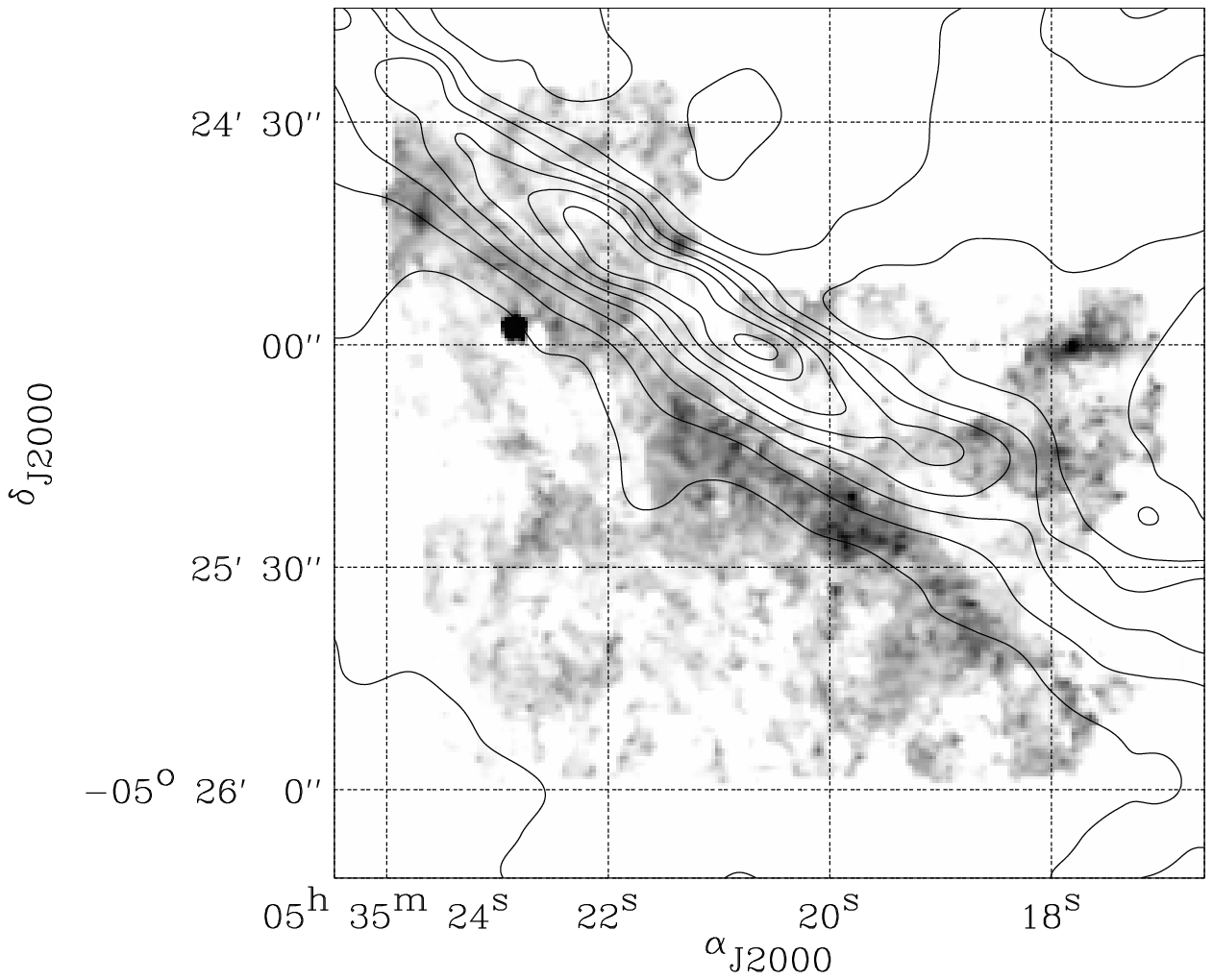,width=8.8cm,angle=00}}
\caption{ The 9.6\mum~ feature (top) and 6.2\mum-AIB (bottom) both in
contours superimposed to the $v=1\rightarrow 0$ S(1) line emission of
molecular hydrogen taken from van der Werf \etal (\cite{vanderwerf})
(grey scale). The contours correspond to integrated band intensities
from 0.02 to 0.11 by steps of 0.01 (top figure) and 0.045 to 0.27 by
steps of 0.025 (bottom figure) in units of \flux.  }
\end{figure}

Due to the low spectral resolution of the CAM-CVF, however, the
9.6\mum~ feature will certainly blend with the S(3) pure rotational
line of molecular hydrogen - if present. To check this we have
compared our 9.6\mum~ map to that of molecular hydrogen in its
fluorescent vibrational line 1$\rightarrow 0$ S(1)
(2.12\mum). Courtesy of P.P. van der Werf (van der Werf \etal
\cite{vanderwerf}), we reproduce in Fig. 9 the map of the fluorescent
molecular hydrogen emission. This latter correlates better with the
AIB emission as traced by the 6.2\mum-feature (bottom figure) than it
does with the tentative crystalline silicate emission (top), namely
they both peak along the bar. This is not surprising because the H$_2$
and AIB emitters require shielding from far-UV radiation to
survive. Conversely, the 9.6\mum~silicate feature is stronger where
H$_2$ is weak as can be seen around $\theta^2$~Ori~A. In addition, the
H$_2$ S(3) rotational line at 9.66\mum~ is detected in the ISO-SWS
spectrum of the Orion bar presented in Verstraete \etal (1999, in
preparation) with an intensity of $6\times 10^{-7}$ W m$^{-2}$
sr$^{-1}$. This value is a factor of 16 below the median flux of the
9.6\mum~ feature in our map, namely $10^{-5}$ W m$^{-2}$ sr$^{-1}$. We
can thus safely conclude that our 9.6\mum-emission predominantly
originate from silicates.
%
A confirmation of the identification of the 9.6\mum~ band with a
crystalline silicate dust component would be possible if a second
signature band were seen in our spectra. The SWS spectrum (Fig. 4)
shows only broad emission bands that are difficult to characterise,
and additionally, the chacteristic crystalline olivine band in the
$11.2-11.4$\mum~ region (\eg Jaeger \etal \cite{Jaeger}), if present,
is blended with the 11.2\mum~ aromatic hydrocarbon
feature. Additionally, most of the chacteristic crystalline bands fall
longward of the CVF spectra. Thus, it is difficult to
self-consistently confirm the 9.6\mum~ band identification with the
presented data.
%

In summary, emission in the 9.7\mum~band of amorphous silicate
emission exists everywhere inside the Orion \HII~ region. Previously,
amorphous silicate emission had only been seen in the direction of the
Trapezium (Stein \& Gillett \cite{Stein}; Forrest \etal
\cite{Forrest}; Gehrz \etal \cite{Gehrz}). We may assume that the
18\mum~ band is also widely present in the region, as witnessed by the
single SWS spectrum (Fig. 4) and by the generally rising long
wavelength end of ISOCAM spectra; the two spectra shown, Figs. 3 and
10 are quite representative of the steeply rising continuum longward
of 15\mum.

\subsection{ The interstellar silicate and H$_2$ emission around 
	     \mbox{$\theta^2$~Ori~A}}

The case of $\theta^2$~Ori~A is particularly interesting because the
geometry is simple and therefore allows quantitative
calculations. Moreover, the thermal radio continuum, the recombination
lines and the fine--structure lines are faint in the neighbourhood of
this star (Felli \etal \cite{Felli}; Pogge \etal \cite{Pogge}; Marconi
\etal \cite{Marconi}, and the present paper, Fig. 2).
$\theta^2$~Ori~A is classified as an O9.5Vpe star and shows emission
lines (see e.g. Weaver \& Torres--Dodgen \cite{Weaver}). It is a
spectroscopic binary and an X-ray source.  There is little gas left
around the star and the observed silicate dust (Fig. 8) is almost all
that is visible of the interstellar material left over after its
formation. Indeed, O stars are not known to produce dust in their
winds which are probably much too hot, so that the silicates we see
here must be of interstellar origin.
 
The mid--IR continuum observed towards $\theta^2$~Ori~A can be
accounted for by combining emission of warm silicate and carbon grains
(see Fig. 10). The
model continuum was obtained in the same way as for the SWS
observation (see Fig. 4) and with the same assumptions. The grain
temperatures are consistent with the heating of interstellar grains by
the strong radiation field of the star.

\begin{figure}[t!]
\vspace{0cm}
{\psfig{figure={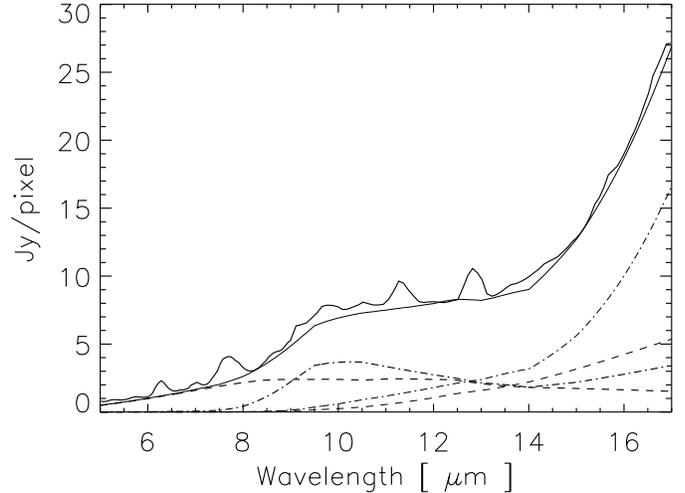},width=8.8cm }}
\caption{CVF spectrum towards $\theta^2$ Ori A (heavy solid line).
The ordinates give fluxes in Jy per $6\arcsec\times6\arcsec$ pixel.
The fit to the continuum is shown by the thin solid line.
The fit (see \S3) to these data comprises, from top to bottom
on the right--hand axis: 100--K amorphous astronomical silicate
(dot-dashed line), 
110--K amorphous carbon (dashed line ),
235--K amorphous astronomical silicate (dot-dashed line) and 
330--K amorphous carbon emission (dashed line). }
\end{figure}

As discussed above, the band near 9.6 $\mu$m (Fig. 6 bottom and
Fig. 8) may be due to crystalline silicates, any contribution of the
S(3) H$_2$ line to this band is minor. Another band at 14 $\mu$m (see
Figs. 4 and 10) might also be due to crystalline silicates. Amongst
the crystalline silicates whose mid--IR absorption spectra are shown
by Jaeger \etal (\cite{Jaeger}), synthetic enstatite (a form of
pyroxene) might perhaps match the $\theta^2$ Ori A spectrum.  The
interest in the possible presence of crystalline silicates around this
star is that they would almost certainly be of interstellar origin,
pre--dating the formation of the star. Observations at longer
wavelengths are needed for a definitive check of the existence of
crystalline silicates and for confirming their nature. Such
observations do not exist in the ISO archives and should be obtained
by a future space telescope facility.

\section{Conclusions}

We obtained a rather complete view of the infrared emission of the
Orion nebula and its interface with the adjacent molecular cloud. The
most interesting results are the observation of amorphous, and
possibly crystalline, silicates in emission over the entire \HII~
region and in an extended region around the bright O9.5Vpe star
$\theta^2$ Ori A. We have fitted the mid--IR continuum of the \HII~
region and around $\theta^2$ Ori A with the emission from amorphous
silicate and amorphous carbon grains at the equilibrium temperatures
predicted for the grains in the given radiation field.  This shows
that both types of grains can survive in the harsh conditions of the
\HII~ region. A number of bands (the 9.6\mum~bump seen in Fig. 6; the
excess 14\mum~emission indicated in Figs. 4 and 10) suggest emission
from crystalline silicates (essentially forsterite) in the \HII~
region.  Crystalline silicates may also exist around $\theta^2$ Ori A,
but further, longer wavelength observations are required to confirm
their presence. \\

Do the observed crystalline silicates result from processing of
amorphous silicates in the \HII~ region or in the environment of
$\theta^2$ Ori A?  Silicate annealing into a crystalline form requires
temperatures of the order of 1000~K for extended periods (Hallenbeck
\etal \cite{Hallenbeck}). The dust temperatures observed in the \HII~
region and around $\theta^2$ Ori A are considerably lower than this
annealing temperature. One might however invoke grain heating
following grain--grain collisions in the shock waves that are likely
to be present in the \HII~ region. However, grain fragmentation rather
than melting is the more likely outcome of such collisions (Jones
\etal \cite{JTH}). It is probable that the crystalline silicates observed here
were already present in the parent molecular cloud, and probably
originate from oxygen--rich red giants. \\

Emission by both amorphous and crystalline silicates has been observed
with ISO around evolved stars (Waters \etal \cite{Waters}; Voors \etal
\cite{Voors}). The crystalline silicates there must have been produced 
locally by annealing of amorphous silicates. Gail \& Sedlmayr
(\cite{Gail}) have shown that this is possible, and that both
amorphous and crystalline forms can be released into the interstellar
medium. However, there is no evidence for absorption by crystalline
silicates in the general interstellar medium in front of the deeply
embedded objects for which amorphous silicate absorption is very
strong (Demyk \etal 1999, Dartois \etal \cite{Dartois}).
Consequently, crystalline silicates represent only a minor fraction
compared to amorphous silicates. It would be difficult to detect the
emission from a small crystalline component of dust in the diffuse
interstellar medium because the dust is too cool (T$\,\sim 20\,$K) to
emit strongly in the $15-40\,\mu$m wavelength region. Observations of
\HII~ regions and bright stars provide the opportunity of observing
this emission due to the strong heating of dust.  Emission from
amorphous and crystalline silicates is seen around young stars
(Waelkens \etal \cite{Waelkens}; Malfait \etal \cite{Malfait}) as well
as in comets (Crovisier \etal \cite{Crovisier}). There are also
silicates in meteorites, but their origin is difficult to determine
because of secondary processing in the solar system. Crystalline
silicates in comets, and perhaps in interplanetary dust particles
believed to come from comets (Bradley \etal
\cite{Bradley}), must be interstellar since the material  in comets never
reached high temperatures. However, the silicates probably experienced changes
during their time in the interstellar medium. It is interesting to note that
while very small grains of carbonaceous material exist, there seem to be no 
very small silicate grains in the interstellar medium (D\'esert \etal 1986).

\begin{acknowledgements}
A.P. Jones is grateful to the Soci\'et\'e de Secours des Amis des Sciences
for funding during the course of this work. We are grateful to P.P. van der
Werf for providing us with his map of molecular hydrogen emission. 
\end{acknowledgements}

\begin{flushleft}
{\bf Appendix A: the mid--IR line emission from the Orion nebula}
\end{flushleft}

We presented in Fig. 2 a map of the studied region in the [\NeIII]
line at 15.5 $\mu$m. Fig. 11 displays the map of the [\ArII] line at
7.0 $\mu$m superimposed on the map of the [\ArIII]~9.0$\mu$m line.
These maps illustrate the ionization structure of the Orion
nebula. The spectral resolution of the CVF does not allow a separation
of the the [\ArII] line at 6.99 $\mu$m from the S(5) pure rotation
line of H$_2$ at 6.91 $\mu$m. However, the bulk of the H$_2$ emission
come from deeper in the molecular cloud than that of [\ArII], \ie more
to the south-west (see Fig. 9) and the contamination by the S(5) line
is probably minor.  The SWS spectrum shown here and that taken towards
the bar (Verstraete \etal 1999, in preparation) in which the [\ArII]
and the H$_2$ S(5) line are well separated from each other, show that
the H$_2$ line is a factor 4 or 5 weaker and hence cannot seriously
contaminate the [\ArII] map.

\begin{figure}[t!]
\vspace{-1cm}
{\psfig{file=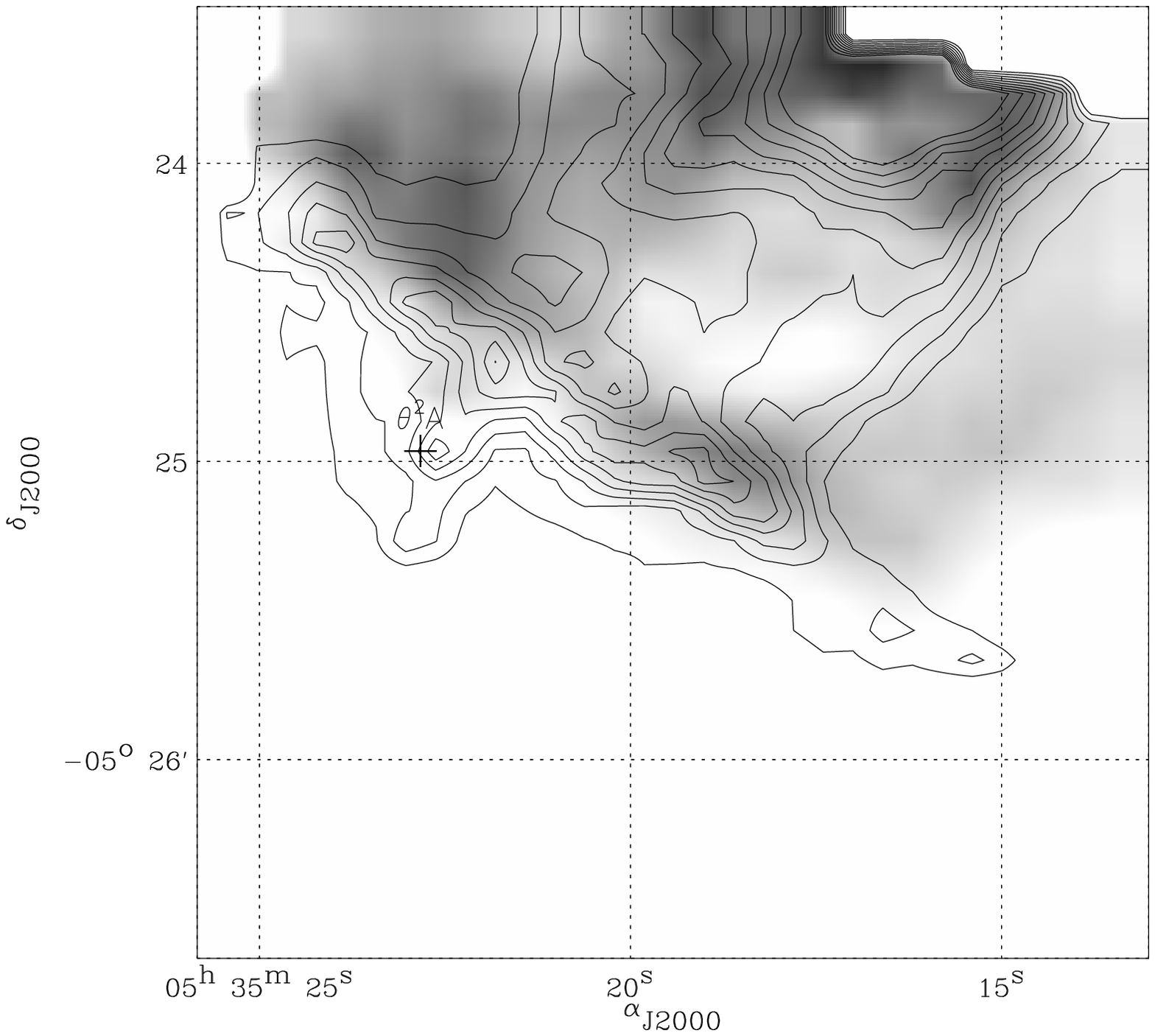,width=8.8cm,angle=00}}
\caption{Map in the line of [\ArII] at 7.0 $\mu$m (contours) superimposed
on the map of the [\ArIII] 9.0 $\mu$m line (grey scale spanning 0.01
to 0.05 \flux). The contours for [ArII] are from 10$^{-3}$ to
10$^{-2}$ \flux~ by steps of 10$^{-3}$. The position of $\theta^2$ Ori
A is indicated by a cross. The peak at 7 $\mu$m at this position is
probably due to the S(5) line of H$_2$ rather than to [\ArII] (see
text).  }
\end{figure} 

The emission by the singly--charged ion [\ArII] is concentrated near
the ionization front on the inner side of the bar. This is very
similar to what is seen in the visible lines of [\NII] $\lambda$6578
and [\SII] $\lambda$6731 (Pogge \etal \cite{Pogge} Fig. $1c$ and
$1d$).  The detailed correspondence between the maps in these three
ions is excellent: note that the optical maps are not much affected by
extinction. The ionization potentials for the formation of these ions
are 15.8, 14.5 and 10.4 eV for \ArII, \NII~, \SII~ respectively, and
are thus not too different from each other.
 
The emission from the doubly--charged ions [\NeIII] and [\ArIII] shows
a very different spatial distribution, with little concentration near
the bar but increasing towards the Trapezium. The [\NeIII] map
(Fig. 2) is very similar to the [\OIII]$\lambda$5007 line map (Pogge
\etal \cite{Pogge} Fig. $1e$), as expected from the similarity of the
ionization potentials of [\NeII] and [\OII], respectively 41.1 and
35.1 eV. However, the distribution of the [\ArIII] line (Fig. 11) is
somewhat different, with a trough where the [\NeIII] and the [\OIII]
lines exhibit maxima. \ArIII~ is ionized to \ArIV~ at 40.9 eV, almost
the same ionization potential as that of \NeII, so that \ArIV~ (not
observable) should co--exist with \NeIII~ and \ArIII~ with \NeII.  A
map (not displayed) in the 12.7 $\mu$m feature, which is a blend of
the 12.7 $\mu$m AIB and of the [\NeII] line at 12.8 $\mu$m, is indeed
qualitatively similar to the [\ArIII] line map in the \HII~ region. It
differs in this region from the maps in the other AIBs, showing that
it is dominated by the [\NeII] line.

As expected, the dereddened distribution of the H$\alpha$ line (Pogge
\etal \cite{Pogge} Fig. $3b$), an indicator of density, is
intermediate between that of the singly--ionized and doubly--ionized
lines.

\begin{flushleft}
{\bf Appendix B: the AIB emission}
\end{flushleft}

Maps of the emission of the 6.2 and 11.3$\mu$m AIBs are shown in
Fig. 12. We do not display the distribution of the other AIBs because
they are very similar.  All the spectra of Figs. 3, 4 and 6 show the
classical UIBs at 6.2, 7.7, 8.6, 11.3 and 12.7 $\mu$m (in the CAM-CVF
data the latter is blended with the [\NeII] line at 12.8
$\mu$m). There are fainter bands at 5.2, 5.6, 11.0,, 13.5 and 14.2\mum
visible in the SWS spectrum of Fig. 3: they may be AIBs as well.  All
the main bands visible in the CVF spectra are strongly concentrated
near the bar. Emission is observed everywhere, because of the
extension of the PDR behind the Orion nebula and the presence of
fainter interfaces to the South--East of the bar. We confirm the
general similarity between the distributions of the different AIBs
through the Orion bar observed by Bregman
\etal  (\cite{Bregman}).

\begin{figure}[t!]
\vspace{-1cm}
{\psfig{file=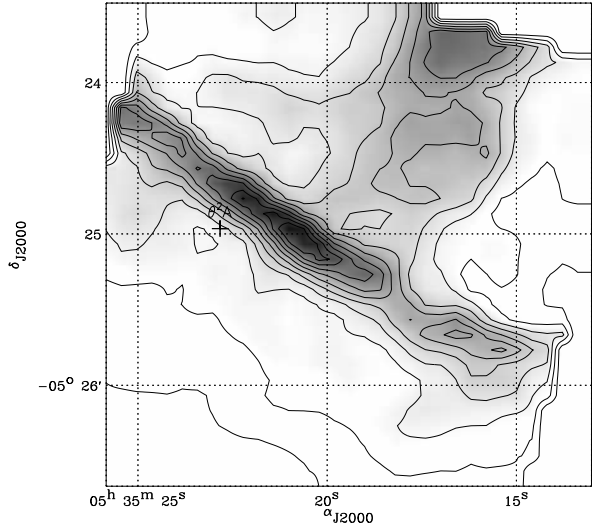,width=8.8cm,angle=00}}
\caption{Map of the 11.3\mum UIB (contours) superimposed on the
6.2\mum UIB map (grey scale from 0.05 to 0.3 \flux). The contours
correspond to integrated band intensities from 0.016 to 0.16 \flux~ by
steps of 0.016. The distributions of the two UIBs are extremely
similar. The position of $\theta^2$ Ori A is indicated by a cross.  }
\end{figure}

We thus conclude that, although the excitation conditions vary greatly
from the Trapezium region towards the South--West of the bar, the
mixing of fore-- and background material along the line of sight does
not allow us to observe spectroscopical changes in the AIB emission
features (due e.g. to ionization or dehydrogenation as in M17-SW,
Verstraete \etal \cite{verstraete}).

\begin{flushleft}
{\bf Appendix C: Estimates of emission strengths}
\end{flushleft}

\begin{figure}[t]
\vspace{-0cm}
{\psfig{file=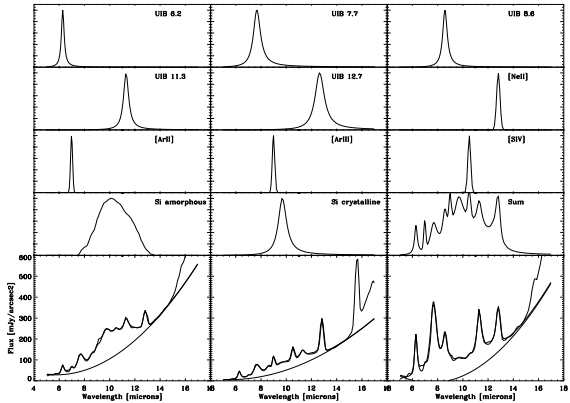,width=8.8cm,angle=00}}
\caption{The eleven line and band templates, normalized to unit peak
intensity, used by the least square fitting code (first eleven panels;
the 12th panel, labeled Sum, shows the combined template). The low
three panels give examples of the fit goodness for three lines of
sight (from left to right): towards $\theta^2$ Ori A; towards a ``hot
spot'' in the \HII~ region; and towards a ``hot spot'' on the AIB
emission. For all lines of sight the fit was stopped at 15\mum~since a
simple parabola could not account for the steep rise at longer
wavelengths.  }
\end{figure}

Spectral emission maps have been obtained using one or another of
three different methods. The emission from well defined and rather
narrow spectral features, viz. AIBs and ions, can be estimated either
by numerical integration of the energy within the line and an ad--hoc
baseline ({\it method 1}), or by simultaneous fit of Lorentz
(Boulanger \etal, \cite{bbcr}) and/or gauss profiles, including a
baseline, determined by a least square fitting algorithm ({\it method
2}). The strength of features not amenable to an analytical
expression, like the suspected amorphous silicate emission (see
Fig. 6) has been estimated using the following method ({\it method
3}). We have constructed an emission template consisting of all the
observed emission features, each one arbitrarily normalised to unit
peak intensity, see Fig. 13. A least square computer code was then
used to obtain, for each of the $32\times 32$ lines of sight, a set of
multiplying coefficients for each feature present in the template plus
a global parabolic baseline so as to minimize the distance between the
model and the data points. The number of free parameters is then
eleven ``line intensities'' and three polynomial coefficients, for a
total of 14 free parameters to be determined from 130 observed
spectral points per line of sight. The main drawback of this method is
that it does not allow for varying line widths or line centres;
however, given the low resolution of ISOCAM's CVF this is not a
serious drawback. We have found that integrated line emission
estimated from methods 2 and 3 give results that agree to within 20
percent; numerical integration of Lorentzian line strengths, on the
other hand, badly underestimates the energy carried in the extended
line widths and hence this method has not been used.


\begin{thebibliography}{}

\bibitem[1996]{allain} Allain T., Leach S., Sedlmayr E., 1996, A\&A 305, 602

\bibitem[1976]{Becklin}
Becklin E.E., Neugebauer G., Beckwith S., et al., 1976, ApJ 207, 770

\bibitem[1998]{bbcr}
Boulanger F., Boissel P., Cesarsky D., Ryter C. 1998, A\&A 339, 194

\bibitem[1992]{Bradley}
Bradley J.P., Humecki H.J., Germani M.S. 1992, ApJ 394, 643 

\bibitem[1989]{Bregman}
Bregman J.D., Allamandola L., Witteborn F.C., Tielens A.G.G.M., Geballe T.R.
1989, ApJ 344, 791

\bibitem[1996a]{CCesarsky}
Cesarsky C.J., Abergel A, Agn\`ese P. \etal ~1996a, A\&A 315, L32

\bibitem[1996b]{M17}
Cesarsky D., Lequeux J., Abergel A., et al., 1996b, A\&A 315, L309


\bibitem[1998]{Coulais}
Coulais A., Abergel A. 1998, 
in ``The Universe as seen by ISO", eds. P. Cox \& M. Kessler, ESA
Special Publications series (SP-427), ESTEC, Noordwijk

\bibitem[1998]{Crovisier}
Crovisier J., Leech K., Bockelee--Morvan D., et al., 1998, 
in ``The Universe as seen by ISO", eds. P. Cox \& M. Kessler, ESA
Special Publications series (SP-427), ESTEC, Noordwijk

\bibitem[1998]{Dartois}
Dartois E., Cox P., Roelfsema P.R. \etal 1998, A\&A 338, L21

\bibitem[1999]{demyk} Demyk K., Jones A.P., Dartois E. \etal 1999, A\&A 349,
267

\bibitem[1986]{Desert86}
D\'esert F.X., Boulanger F., L\'eger A. \etal, 1986, A\&A 159, 328

\bibitem[1990]{Desert90}
D\'esert F.X., Boulanger F., Puget J.L., 1990, A\&A 237, 215

\bibitem[1984]{DandL84}
Draine B.T., Lee H.M. 1984, ApJ 285, 89

\bibitem[1985]{Draine85}
Draine, B.T. 1985, ApJS 57, 587

\bibitem[1993]{Felli}
Felli M., Churchwell E., Wilson T.L., Taylor G.B. 1993, A\&AS 98, 137

\bibitem[1975]{Forrest}
Forrest W.J., Gillett F.C., Stein W.A. 1975, ApJ 195, 423

\bibitem[1999]{Gail}
Gail H.-P., Sedlmayr E. 1999, A\&A 347, 594

\bibitem[1975]{Gehrz}
Gehrz R.D., Hackwell, J.A., Smith, J.R. 1975, ApJ 202, L33


\bibitem[1998]{Hallenbeck}
Hallenbeck S.L., Nuth J.A., Daukantas P.L. 1998, Icarus 131, 198



\bibitem[1998]{Jaeger}
Jaeger C., Molster F.J., Dorschner J., et al., A\&A 339, 904

\bibitem[1996]{JTH}
Jones A.P., Tielens A.G.G.M., Hollenbach D.J. 1996, ApJ 469, 740

\bibitem[1998]{Jones}
Jones A.P., Frey V., Verstraete L., Cox P., Demyk K. 1998,
in ``The Universe as seen by ISO", eds. P. Cox \& M. Kessler, ESA
Special Publications series (SP-427), ESTEC, Noordwijk

\bibitem[1993]{Koike}
Koike C., Shibai H., Tuchiyama A. 1993, MNRAS 264, 654


\bibitem[1998]{Malfait}
Malfait K., Waelkens C., Waters L.B.F.M., et al., 1998, A\&A 332, L25

\bibitem[1998]{Marconi}
Marconi A., Testi L., Natta A., Walmsley C.M. 1998, A\&A 330, 696

\bibitem[1973]{Ney}
Ney E., Strecker D., Gehrz R. 1973, ApJ 180, 809

\bibitem[1992]{Pogge}
Pogge R.W., Owen J.M., Atwood B. 1992, ApJ 399, 147

\bibitem[1999]{koko}
Okumura K. 1999, The ISO point spread function and CAM beam profiles,
   proceedings ``ISO beyond point sources''

\bibitem[1991]{Rouleau}
Rouleau F., Martin P.G. 1991, ApJ 377, 526

\bibitem[1998]{Starck}
Starck J.L., Abergel A., Aussel H., et al., 1998, A\&AS 134, 135

\bibitem[1969]{Stein}
Stein W.A., Gillett F.C. 1969, ApJ 155, L197



\bibitem[2000]{vanKerckhoven}
van Kerckhoven, C., Hony, S., Peters, E., Tielens A.G.G.M. 2000,
in ``ISO beynd the peaks: The 2nd workshop on analytical spectroscopy'',
in press

\bibitem[1996]{vanderwerf}
van der Werf P., Stutzki J., Sternberg A., Krabbe A. 1996, A\&A 313, 633

\bibitem[1996]{verstraete} 
Verstraete L., Puget J.L., Falgarone E. et al., 1996, A\&A 315, L337

\bibitem[1998]{Voors}
Voors R.H.M., Waters L.B.F.M., Morris P.W. et al., 1998, A\&A 341, L193

\bibitem[1996]{Waelkens}
Waelkens C., Waters L.B.F.M., de Graauw M.S. \etal 1996, A\&A 315, L245

\bibitem[1996]{Waters}
Waters L.B.F.M., Molster F.J., de Jong T. \etal 1996, A\&A 315, L361

\bibitem[1997]{Weaver}
Weaver W.B., Torres--Dodgen A.V. 1997, ApJ 487, 847

\end{thebibliography}
\end{document}